\documentclass[conference]{IEEEtran}
\IEEEoverridecommandlockouts
\usepackage{cite}
\usepackage{amsmath,amssymb,amsfonts}
\usepackage{algorithmic}
\usepackage{graphicx}
\usepackage{textcomp}
\def\BibTeX{{\rm B\kern-.05em{\sc i\kern-.025em b}\kern-.08em
    T\kern-.1667em\lower.7ex\hbox{E}\kern-.125emX}}
\begin{document}

\title{Decentralised firewall for malware detection\\}

\author{\IEEEauthorblockN{Saurabh Raje}
\IEEEauthorblockA{\textit{Department of Computer Science and Information Systems} \\
\textit{Birla Institute of Technology and Science}\\
Pilani, India \\
saurabh.mraje@gmail.com}
\and
\IEEEauthorblockN{Shyamal Vaderia}
\IEEEauthorblockA{\textit{Department of Computer Science and Information Systems} \\
\textit{Birla Institute of Technology and Science}\\
Pilani, India \\
vaderiashyamal@gmail.com}
\and
\IEEEauthorblockN{Neil Wilson}
\IEEEauthorblockA{\textit{Department of Computer Science and Information Systems} \\
\textit{Birla Institute of Technology and Science}\\
Pilani, India \\
neilvpwilson@gmail.com}
\and
\IEEEauthorblockN{Rudrakh Panigrahi}
\IEEEauthorblockA{\textit{Department of Electrical and Electronics Engineering} \\
\textit{Birla Institute of Technology and Science}\\
Pilani, India \\
rudrakh97@gmail.com}
}

\maketitle

\begin{abstract}
This paper describes the design and development of a decentralized firewall system powered by a novel malware detection engine. The firewall is built using blockchain technology. The detection engine aims to classify Portable Executable (PE) files as malicious or benign. File classification is carried out using a deep belief neural network (DBN) as the detection engine. Our approach is to model the files as grayscale images and use the DBN to classify those images into the aforementioned two classes. An extensive data set of 10,000 files is used to train the DBN. Validation is carried out using 4,000 files previously unexposed to the network. The final result of whether to allow or block a file is obtained by arriving at a proof of work based consensus in the blockchain network.
\end{abstract}


\section{Introduction}
This study was motivated by the fact that the current malware detection systems are more predictable (in their behavior) than predictive. Moreover, there is great potential in leveraging the increased security offered by blockchain technology in securing local networks. The conventional strategy for malware detection is to maintain a database of malicious signatures. A similar signature is generated for any incoming file and is run through the database. The system heavily relies on updates to this database for currency. There is hence a dire need for malware detection that can perform well on previously unseen attacks. This has the potential to outperform human prediction in terms of time and versatility. Therefore, it is imperative that a non-conventional and scalable approach be followed while modeling malware. \par
Modeling the malware as a grayscale image (as per our model) scales the complexity of the problem as per the accuracy required and computational power available. The blockchain based architecture ensures that no single node can compromise the network. In order to overpower the network, a user needs to have control of more than 50\% nodes in the network. The proof of work associated with a blockchain requires that the computational power needed is exponential in the length of the current longest chain. Furthermore, the final consensus also takes into account the ‘trust value’ of each node in the network. This enables us to suppress the contribution of defective nodes. \par
Several methods have been proposed for intelligent malware detection systems, e.g., a faster update and maintenance system for decentralized anti-virus software using distributed blockchain network and feedforward scanning \cite{b1}, a deep learning based method for automatic malware signature generation and classification \cite{b2}, a deep learning approach using self-taught learning (STL) for Network Intrusion Detection System \cite{b3}. This paper uses a unique amalgamation of blockchain technology and deep learning in order to design a full-proof heuristic solution. \par
The rest of this article is organized as follows. Section 2 of this paper describes the system design and program flow, while section 3 describes the novel detection engine that uses DBN. Section 4 contains the results obtained for the malware detection. Finally, section 5 concludes this paper and hence the study.

\section{System Design And Flow}
The task of malware detection is carried out by the detection engine that will be treated as a black box for the scope of this section. 
The crux of the solution lies in utilising all the machines in a local network for every incoming file to be tested. \par
Whenever a node is added to the network, a central server ships an unique trained model to it that will serve as the detection engine at that node. Any incoming file to a node in the network is first broadcasted to all the nodes (machines) in the local network on which this system is deployed. This broadcast will not be carried out in the blockchain. At every node, this file is run through the detection engine (trained model) present at that node. A point to be noted here, is that all the nodes possess unique detection engines. The numeric result produced by an engine represents the probability of the file being malicious. This is hashed by the node’s own key and added to the blockchain as a transaction. \par
\begin{figure}
\centerline{\includegraphics[width=250pt]{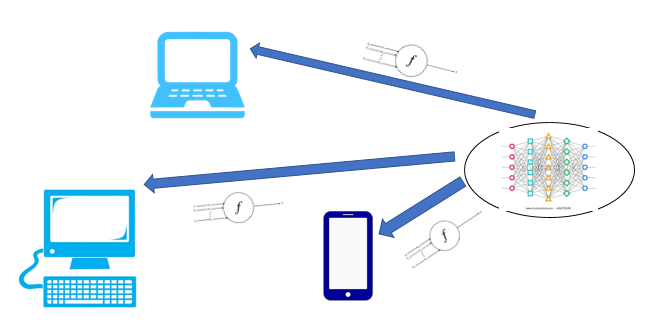}}
\caption{Uniquely trained model shipped (by a central server) to each node in the P2P network.}
\label{fig1}
\end{figure}
\begin{figure}
\centerline{\includegraphics[width=250pt]{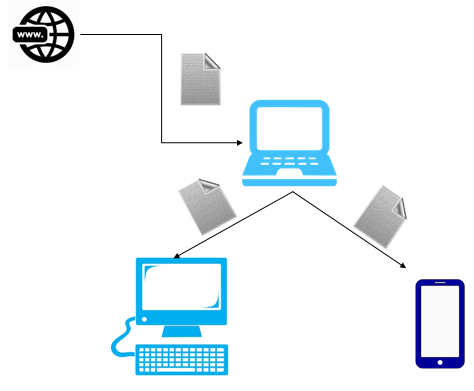}}
\caption{File downloaded from the internet is broadcasted in the P2P network.}
\label{fig2}
\end{figure}
\begin{figure}
\centerline{\includegraphics[width=250pt]{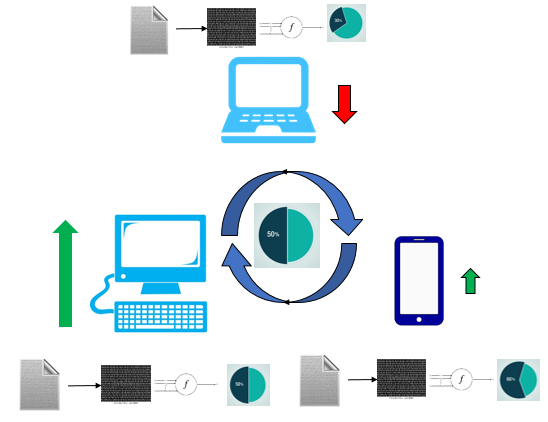}}
\caption{Each machine processes the same file and determines the probability of the file being malicious. The network finally comes to a consensus.}
\label{fig3}
\end{figure}

The node that broadcasted the file can then traverse the chain to get the final verdict by performing a weighted average of the probability values added to the chain by other nodes in the network, corresponding to the same file. The weight given to each node’s probability value is a direct measure of the trust of that node in the network. This trust is updated after every transaction based on the deviation of the node’s result from the average probability over all the nodes in the network. Fig. \ref{fig1}, Fig. \ref{fig2} and Fig. \ref{fig3} illustrate this flow.

Any unauthorized modifications to the blockchain in a node will be automatically invalidated as the hash of that block and the subsequent ones will be different. Hence the results given out by the nodes cannot be tampered with. We have used the ethereum network to implement our blockchain. Ethereum is an open source, public, blockchain-based platform which provides smart contract functionality. These smart contracts are executed by the ethereum virtual machine(EVM). Smart contracts are self-executing, self-enforcing protocols that facilitate the enforcement of predefined contracts (event responses) in a blockchain. We have used solidity language for implementing ethereum smart contracts. \par
We implemented the firewall on a testnet using TestRPC (a nodeJS based client that simulates the Ethereum blockchain network). TestRPC creates a local P2P network consisting of 10 accounts. Each node in the network controls one of these accounts. Using smart contracts these accounts come to a consensus on the nature of a file.  All the events taking place in the network like downloading of an external file, running the neural net, trust updating etc are hashed and stored in the blockchain as transactions (also called ledgers).

\section{Detection Engine}
The malware detection problem has effectively been reduced to an image classification problem by modeling the PE files as grayscale images.  Every file is read byte by byte and each byte value is stored as the grayscale intensity for the corresponding pixel. Since files can easily exceed few megabytes in size, the corresponding image will be of the order of megapixels. There is hence a need to downscale the images to a fixed number of pixels corresponding to the number of input neurons. This process is carried out using python scripts that use the pillow library for image processing. These images provide the system with the insights from the headers and the machine code of the file, in a different form. \par
It is important to note that this differs slightly from the image classification problem with respect to translational and rotational invariance. Since the spatial distribution of features in the grayscale image represents the order of the corresponding program elements of the PE file, it implies that a change in the spatial distribution indicates a change in the sequence of the program and can not be neglected. This is one reason why a convolutional neural network (CNN) (or a related architecture) has not been considered. It would fundamentally distort the problem because CNNs exploit translational and rotational invariance in the image for increasing the detection accuracy. \par
Since the data is noisy, it is better to use a generative model for gathering hidden feature representations that could aid in the classification problem. Using DBN to detect malware has been attempted previously, but with an entirely different way of modeling the data \cite{b4},\cite{b5}. \par
The dataset used in this study consists of 9342 malicious files across 25 malware families and 4986 known benign files. The malicious dataset was obtained from the MALIMG dataset \cite{b6}. The benign files were obtained from vanilla windows installations. The malicious dataset consisted of grayscale images. The benign files were converted byte by byte into a grayscale image and downscaled it to 64X64. We now describe the detailed neural network architecture and training algorithm.

\subsection{Overview of DBN}\label{AA}
The Deep Belief Network (DBN) is a generative model that consists of multiple stacked Restricted Boltzmann Machines (RBM). RBM is a type of unsupervised neural network that can represent non-linear patterns in data. An RBM is a two-layer model that consists of a single visible layer and a single hidden layer. Neurons of an RBM are stochastic binary units, hence they can be in only one of the two states, either ‘on’ or ‘off’, with a given probability.
\begin{figure}
\centerline{\includegraphics[width=250pt]{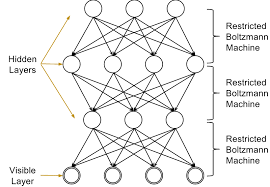}}
\caption{DBN as a set of stacked RBMs.}
\label{fig4}
\end{figure}
Two layers of RBM are fully connected via symmetric and undirected weights, and neurons within a layer are not connected (shown in Fig 4). In an RBM, each set of weight and biases defines energies of joint configurations of visible and hidden units $ E \left( v,h \right)$, which is defined in \eqref{1}.

\begin{equation}
\left( v,h \right) = \sum _{i \in vis}^{}v_{i}b_{i}~+ \sum _{k \in hid}^{}h_{k}b_{k}+ \sum _{i,j}^{}v_{i}h_{k}w_{ik}\label{1}
\end{equation}

In \eqref{1} $ v $, $ h $ represents the configuration of hidden and visible unit respectively. $ v_{i} $, $ j_{i} $ are binary state of the visible unit $ i $ and hidden unit $ j $.

\begin{equation}
p \left( v,h \right) ~=~\frac{e^{-E \left( v,h \right) }}{ \sum _{u,g}^{}e^{-E \left( u,g \right) }}\label{2}
\end{equation}

\begin{equation}
p \left( v,h \right) ~~ \propto ~e^{-E \left( v,h \right) }~\label{3}
\end{equation}

As shown in \eqref{2}, The probability of the network stabilising to a particular joint configuration of visible and hidden units depends on the energy of that configuration compared with the energy of all other joint configurations. If there are more than few nodes, it’s not computationally possible to compute the normalization term (partition function) in \eqref{2} because it has exponentially many terms. Since as shown in \eqref{3} $ p \left( v,h \right)  $
is directly proportional to $ e^{-E \left( v,h \right) } $
, we use Markov Chain Monte Carlo to get samples from model starting from a global configuration \cite{b7}.

\begin{equation}
p \left( v \right) = \sum _{h}^{}p \left( h \right) ~p \left( v~ \vert ~h \right)\label{4}
\end{equation}

We can express the RBM model as shown in \eqref{4}. If we want to improve $ p \left( v \right)  $, it will suffice to improve $ p \left( h \right)  $. To improve $ p \left( h \right)  $, we need it to be a better model than $p \left( h,W \right)$. We hence add another layer to model $ p \left( h \right)  $ as its input layer in order to improve the hidden feature representation. This recursive logic can be applied as many times in order to get the network deeper by adding more layers. This intuitively describes why stacking RBMs will make a better learning architecture. This is called a DBN. Experimentally, for this data, 2 such layers have sufficed to give the best accuracy as shown.

\subsection{Training procedure}
To train a DBN for obtaining hidden representations from some data, one must consider it as a set of multiple stacked RBMs. The RBMs are then trained separately. Layer by layer greedy pretraining is done bottom up from the RBM that has the input layer up to the last one preceding the softmax output layer. This requires us to consider the input of one RBM as the output of the one right below it (Fig. 4). \par
To train RBM following procedure is used. \eqref{5} shows derivative of $ log $
of probability of visible vector $ v $
with respect to weights. The angular brackets indicates expected value of the given quantity. $ 〈v_{i}h_{j}〉_{data} $
represents expected value of training data when visible vector $ v $
is fixed and equilibrium is attained as per the energy function described in the previous section. $ 〈v_{i}h_{j}〉_{model}~ $
represents the expected value of model when it attains equilibrium and all units are allowed to change. 

\begin{equation}
\frac{ \partial log~P \left( v \right) }{ \partial w_{ij}}=〈v_{i}h_{j}〉_{data}~-〈v_{i}h_{j}〉_{model}~ \label{5}
\end{equation}

Now we can show that the update rule for the weights will be as shown in Eq. (6), where $  \varepsilon  $ is the learning rate.

\begin{equation}
\Delta w_{ij}= \varepsilon ~ \left( 〈v_{i}h_{j}〉_{data}~-〈v_{i}h_{j}〉_{model}~ \right) \label{6}
\end{equation}

To calculate the weight update shown in \eqref{6} we use the contrastive divergence algorithm by alternatively sampling hidden units and visible units from one another \cite{b8}. In the sampling process, binary states of visible units are calculated using \eqref{7} and binary states of hidden units are calculated using \eqref{8}, where $ sigm $ represents the sigmoid function. In theory, the sampling should be done an infinite number of times. \par
However, it is extremely inefficient to do so and hence in practice we use contrastive divergence (CD) algorithm (explained below) in which sampling is done for only a few iterations.~If we start sampling the hidden units from the  data on the visible units and then the visible units from the state of the hidden units, the network configuration slowly enters the regions of equilibrium defined as per the energy function of the current weights and biases. When we know weights are inaccurate, it’s waste of time to let it go all the way to attain equilibrium in those regions. The final objective is to raise the probability of data being generated at the visible units. We hence change the weights right after a few (finite) iterations of the sampling process such that the regions of equilibrium shall, with a higher probability, contain the data (or a reasonable approximation thereof) on the visible units. Hence, a good compromise between speed and correctness is to start with small weights and use CD for one iteration, and as weights grow, to increase the number of iterations. In our design, the number of iterations (k) for CD is modeled as shown in \eqref{9}, where $ ep $
represents the number of epochs for which network is being trained.

\begin{equation}
p \left( v_{i}=1~ \vert ~h \right) ~=~sigm \left( ~-b_{i}- \Sigma _{j}w_{ij}h_{j} \right)\label{7}
\end{equation}

\begin{equation}
p \left( h_{j}=1~ \vert ~v \right) ~=~sigm \left( ~-c_{j}- \Sigma _{i}w_{ji}v_{i} \right)\label{8}
\end{equation}

This complete pretraining procedure, however, is not sufficient. It is necessary to carry out fine tuning with some labeled data, in order to make the class boundaries in the probability density function more distinct. The weight updates in this phase are small in magnitude but result in a much better accuracy. This phase entails backpropagating the error from the distribution of the output softmax layer to the RBMs. It also makes the network scale up better with respect to the number of layers.

\begin{equation}
k=~\frac{ep}{10}~+1\label{9}
\end{equation}

\begin{equation}
\varepsilon ~=~\frac{1}{1+e^{ \left( \frac{ep}{10}-5l \right) }}\label{10}
\end{equation}

The fine-tuning process (as per our design) is carried out by using stochastic gradient descent on the output softmax layer with the learning rate modeled as a function of the epochs and layer as shown in \eqref{10}.

\subsection{Choosing hyperparameters}
While there are some notions on how to optimise the hyperparameters of a DBN, it finally comes down to ample experimentation to find the right set of parameters. These include the number of neurons per layer and number of layers at the outset. A study was carried out to investigate the correlation between the number of neurons in a layer to the accuracy obtained for a classification problem \cite{b9}. We hence use a deep belief network with 3000 hidden units in two hidden layers each. Since the images have a downscaled resolution of 64X64, the input layer has 4096 neurons. The output is a softmax layer with two neurons since it is a two class problem. The learning and contrastive divergence rate is not constant and is described above. Training is done in mini-batches of size 10.

\section{Results}

We have done our experimentation on two different architectures of DBN, $ DBN^{2} $
 and $ DBN^{3} $
. $ DBN^{2} $
contains two stacked RBMs and $ DBN^{3} $
contains three stacked RBMs.  Each layer of $ DBN^{2} $
and $ DBN^{3} $
contains 3000 hidden unit. Results of our experiment, for hyperparameters defined in the previous section, is shown in TABLE I. Since $ DBN^{2} $
performed better than $ DBN^{3} $
, we decided not to experiment on $ DBN^{4} $
, i.e. a DBN with 4 layers.

\begin{table}
\caption{Accuracy and TPR}
\begin{center}
\begin{tabular}{|c|c|c|}
\hline
Architecture & Accuracy & TPR  \\
\hline
$ DBN^{2} $& 89.28\% & 0.9826   \\
\hline
$ DBN^{3} $& 88.14\%& 0.9789  \\
\hline
\end{tabular}
\label{tab1}
\end{center}
\end{table}

\section{Conclusion}
This study is a proof of concept for the automation of heuristic malware detection by which unknown malicious attacks may also be prevented. It also highlights the importance and feasibility of amalgamation of various technologies for cybersecurity. The integration of blockchain technology with the novel detection engine provides full security over a variably sized network. Future research may be carried out to scale this up in terms of the dimensionality of the data given to the DBN. This might offer more insight into the behavior of the file but will require more computational resources.

\section*{Acknowledgment}
The authors wish to acknowledge the support of Cyberdome, Kerala Police, India for their guidance and domain knowledge pertaining to various malicious attacks and network intrusion threats.  Furthermore, this project would not have been complete without the infrastructure offered by InfinityLabs, UST Global.

\end{document}